\documentclass{article}

\usepackage{amsmath}  
\usepackage{cite}
\usepackage{braket}

\title{\bf{ Two signs of a Schwinger term in a commutator of fermionic currents}}
  \author{J.S.Bhattacharyya\\
 Kanchrapara College,\\
   Kanchrapara,
  743145,\\
India\\}
\date{\today}
\begin{document}

\maketitle
  \abstract{We consider central extensions of two dimensional abelian current algebra and Virasoro algebra and see that the sign of the Schwinger term changes if we arrange the Fourier modes of the fermion in the commutator in the order opposite to the normal. We argue that this is permissible in the case of fermions, but not in the case of bosons.} 

\par
\noindent


\newpage
\numberwithin{equation}{section}  
\section{Introduction}
\par
Careful computation of the commutators of fermionic currents in a quantum field theory reveals that they do not always have the form anticipated from naive manipulations. Additional terms usually called the Schwinger terms (ST) are to be expected in all current algebras \cite{schw,goto,jack,jordan}
\par
Though apparently incompatible results were reported sometimes \cite{jo,fad,niemi,kob,sonoda,fad2,hosono,stone,tsutsui,kenhof,adam }, the existence of two solutions for the ST different in sign only, seems to be a distinct possibility.
\par
These STs arise due to the short distance singularities of the current - current correlation functions and can be computed in many ways as discussed in the literature. The oldest among them is the canonical method. This is what we want to pursue in this letter.

\section{Abelian Current Algebra}
\par
For simplicity we first consider the 1+1 dimensional model described by the action
\begin{eqnarray}\label{S}
S&=&\frac{i}{2\pi}\int_0^\pi d\sigma\int_{-\infty}^\infty d\tau \bar{\psi}\gamma^\mu\partial_\mu\psi \\\nonumber
&=&\int d ^2\sigma (\psi_+^\dagger\partial_-\psi_++\psi_-^\dagger\partial_+\psi_-)
\end{eqnarray} 
and see that the ambiguities in the sign of the ST as mentioned earlier are actually consequences of those in ordering the operators in the current - current commutator as appropriate for the states they are applied to \cite{isler,vladimir,adam2}. Here the two component spinor

 \begin{equation}
 \psi=\begin{pmatrix}\psi_- \\\psi_+ \end{pmatrix}
\end{equation}
 \begin{eqnarray}
\gamma^0&=&\begin{pmatrix}0&-i\\i&0 \end{pmatrix}\\\nonumber
\gamma^1&=&\begin{pmatrix}0&i\\i&0 \end{pmatrix}
\end{eqnarray} 
and $\sigma_\pm=\tau\pm\sigma$ are the light cone variables.
\par
So from the equation of motions $\partial_\pm\psi_\mp=0$ we get for the left moving piece for example
 \begin{eqnarray}
\psi_+ &=\sum b_k e^{-ik(\tau+\sigma)}\\\nonumber
\psi^\dagger_+ &=\sum\bar{b}_k e^{-ik(\tau+\sigma)}
\end{eqnarray} 
where $ \bar{b}_{-k}=b^\dagger_k $.
\par
The canonical anti-commutation relations yield
\begin{eqnarray}
\{b_k,\bar {b}_{k^\prime}\}&=&\delta_{k+k^\prime}\\\nonumber
\{b_k,b_{k^\prime}\}&=&\{\bar{b}_k,\bar{b}_{k^\prime}\}\\\nonumber
&=&0
\end{eqnarray} 
Here $k$ can assume both integral and half-integral values \cite{ram,nev}.
We consider the $U(1)$ current algebra  \cite{manton}  to start with. The current is
\begin{equation}
j^\mu=\bar{\psi}\gamma^\mu\psi
\end{equation}
Hence,

\begin{equation}\label{c}
j_+=\psi^\dagger_+\psi_+
\end{equation}

with the modes
\begin{equation}
T_k=\sum\bar{b}_{k-q}b_q
\end{equation}
 
The anomalous commutator
\begin{eqnarray}
[T_k,T_{-k}]&=&\sum[\bar{b}_{k-q}b_q,\bar{b}_{-k-q^\prime}b_{q^\prime}]\\\nonumber
&=&\sum(\bar{b}_{k-q}\{b_q,\bar{b}_{-k-q^\prime}\}b_{q^\prime}-\bar{b}_{-k-q^\prime}\{\bar{b}_{k-q},b_{q^\prime}\}b_q)\\\nonumber
&=&\sum[\bar{b}_{k-q}b_{q^\prime}\delta_{q-k-q^\prime}-\bar{b}_{-k-q^\prime} b_q\delta_{k-q+q^\prime}]\\\nonumber
&=&\sum(\bar{b}_{k-q}b_{q-k}-\bar{b}_{-q}b_q)
\end{eqnarray}
Thus,
\begin{eqnarray}\label{n}
[T_k,T_{-k}]&=&:[T_k,T_{-k}]:+\sum_{q<k}1-\sum_{q<0}1\\\nonumber
&=&:[T_k,T_{-k}]:+k
\end{eqnarray}
 
where $ :[T_k,T_{-k}]:=\sum:(\bar{b}_{k-q}b_{q-k}-\bar{b}_{-q}b_q):=0 $. 
\par
Alternatively, we can place the annihilation operators to the left of the creation operators, we call it  anti-normal ordering and use the symbol $::$ $::$ for it. Thus,
\begin{eqnarray}\label{a}
[T_k,T_{-k}]&=&::[T_k,T_{-k}]::+\sum_{q>k}1-\sum_{q>0}1\\\nonumber
&=&::[T_k,T_{-k}]::-k
\end{eqnarray}
where $ ::[T_k,T_{-k}]::=\sum::(\bar{b}_{k-q}b_{q-k}-\bar{b}_{-q}b_q)::=0$. Thus 
\begin{equation}\label{A}
[T_k,T_{-k}]=\pm k
\end{equation}
\par
It is common practice to calculate anomaly by taking expectation values of the commutator in a suitable state. For example in Feynman propagator computations we take the vacuum expectation value. In CFT also we tacitly do the same for radial ordering. In both cases we get $+k$ for the ST. It can be checked from the following equation:
\begin{eqnarray}\label{nn}
&&\bra{0}[T_1,T_{-1}]\ket{0}\\\nonumber
&&=\bra{0}T_1T_{-1}\ket{0}\\\nonumber
&&=\sum_{q,q^\prime}\bra{0}\bar{b}_{1-q}b_q\bar{b}_{-1-q^\prime}b_{q^\prime}\ket{0}\\\nonumber
&&=\bra{0}\bar{b}_\frac{1}{2}b_\frac{1}{2}\bar{b}_{-\frac{1}{2}}b_{-\frac{1}{2}}\ket{0}\\\nonumber
&&=\bra{0}\bar{b}_\frac{1}{2}b_{-\frac{1}{2}}\ket{0}\\\nonumber
&&=1
\end{eqnarray} 
\par
It is easy to check that we get the other solution by taking the expectation value of the commutator in the completly filled state  $\ket{1}$ that is annihilated by all the creation operators.
\begin{eqnarray}\label{aa}
&&\bra{1}[T_1,T_{-1}]\ket{1}\\\nonumber
&&=-\bra{1}T_{-1}T_1\ket{1}\\\nonumber
&&=-\sum_{q,q^\prime}\bra{1}\bar{b}_{-1-q^\prime}b_{q^\prime}\bar{b}_{1-q}b_q\ket{1}\\\nonumber
&&=-\bra{1}\bar{b}_{-\frac{1}{2}}b_{-\frac{1}{2}}\bar{b}_\frac{1}{2}b_\frac{1}{2}\ket{1}\\\nonumber
&&=-\bra{1}\bar{b}_{-\frac{1}{2}}\bar{b}_\frac{1}{2}\ket{1}\\\nonumber
&&=-1
\end{eqnarray}
It is true for integral modes also.
The last equation is reminiscent of the GNS construction \cite{gel} and Tomonaga states \cite{tomo} and is in conformity with the observations made by the authors of  \cite{stone,isler,vladimir,adam2}. 
\par
Though the Schwinger term changes sign, calculation of the leading short-distance singularities in the OPE of two generators by taking expectation values in either of the states $\ket{0}$ and $\ket{1}$  yield the same result: 
\begin{eqnarray}
&&\bra{1} T_{-k}T_k\ket{1}\\\nonumber
&=&-\bra{1}[T_k,T_{-k}]\ket{1}\\\nonumber
&=&k\\\nonumber
&=&\bra{0}[T_k,T_{-k}]\ket{0}\\\nonumber
&=&\bra{0} T_{k}T_{-k}\ket{0}
\end{eqnarray}
where k is positive.

\par
\section{Virasoro Anomaly}
\par
To have a better understanding of the results, we should consider an algebra like a central extension of the Virasoro algebra \cite{pol}, so that unlike the previous case the operator in the (anti-)normal ordered expression for the commutator does not vanish identically. 
\par
We consider Virasoro algebra with Majorana spinors. So \eqref{S} reduces to 
 \begin{equation}
 S=\int d ^2\sigma (\psi_+\partial_-\psi_++\psi_-\partial_+\psi_-)
\end{equation}
\par
The equations of motion  yield
 \begin{equation}
\psi_+ =\sum b_k e^{-ik(\tau+\sigma)}
\end{equation}
Where
\begin{equation}
\{b_k,b_{k^\prime}\}=\delta_{k+k^\prime}
\end{equation}
\par
So the Fourier modes of the holomorphic component of the energy momentum tensor
\begin{equation}
T_{++}=\psi_+\partial_+\psi_+
\end{equation}
are given by
\begin{equation}
L_k=\sum q b_{k-q}b_q
\end{equation}
Thus
\begin{equation}\label{k-k}
[L_k,L_{-k}]=\frac{1}{4}\sum qq^\prime[b_{k-q}b_q,b_{-k-q^\prime}b_{q^\prime}]
\end{equation}
Now,
\begin{eqnarray}
&&[b_{k-q}b_q,b_{-k-q^\prime}b_{q^\prime}]\\\nonumber
&=&b_{k-q}\{b_q,b_{-k-q^\prime}\}b_{q^\prime}+b_{-k-q^\prime}b_{k-q}\{b_q,b_{q^\prime}\}\\\nonumber
&-&\{b_{k-q},b_{-k-q^\prime}\}b_q b_{q^\prime}-b_{-k-q^\prime}b_q\{b_{k-q},b_{q^\prime}\}\\\nonumber
&=&b_{k-q}b_{q^\prime}\delta_{q-k-q^\prime}+b_{-k-q^\prime}b_{k-q}\delta_{q+q^\prime}\\\nonumber
&-&b_qb_{q^\prime}\delta_{q+q^\prime}-b_{-k-q^\prime}b_q\delta_{q-k-q^\prime}
\end{eqnarray}

So from \eqref{k-k}
\begin{eqnarray}
[L_k,L_{-k}]&=&\frac{1}{4}\sum[q(q-k)(b_{k-q}b_{q-k}-b_{-q}b_q)\\\nonumber
&+&q^2(b_qb_{-q}-b_{q-k}b_{k-q})]
\end{eqnarray}

We write it as
\begin{eqnarray}\label{nnn}
[L_k,L_{-k}]&=&:[L_k,L_{-k}]:+A(k)\\\nonumber
&=&2k:L_0:+A(k)
\end{eqnarray}
where the Virasoro anomaly
\begin{eqnarray}
A(k)&=&\frac{1}{4}[(\sum_{q<k}-\sum_{q<0})q(q-k)\\\nonumber
&+&(\sum_{q>0}-\sum_{q>k})q^2]
\end{eqnarray}
Assuming that q is half integral
\begin{eqnarray}\label{unprim}
A(k)&=&\frac{1}{4}\sum_{q=\frac{1}{2}}^{k-\frac{1}{2}}q(2q-k)\\\nonumber
&=&\frac{1}{4}\sum_{n=1}^k(n-\frac{1}{2})(2n-1-k)\\\nonumber
&=&\frac{k^3-k}{24}
\end{eqnarray}
\par
We can also write
\begin{eqnarray}\label{aaa}
[L_k,L_{-k}]&=&::[L_k,L_{-k}]::+A^\prime(k)\\\nonumber
&=&2k::L_0::+A^\prime(k)
\end{eqnarray}
where
\begin{eqnarray}\label{prim}
A^\prime(k)&=&\frac{1}{4}[(\sum_{q>k}-\sum_{q>0})q(q-k)\\\nonumber
&+&(\sum_{q<0}-\sum_{q<k})q^2]\\\nonumber
&=&-\frac{k^3-k}{24}
\end{eqnarray}

From \eqref{unprim} and \eqref{prim} we see that $A^\prime(k)=-A(k)$. The same is true for integral modes also.

\par
If we want the usual solution $A(k)$ for the anomaly we should take expectation value of the commutator \eqref{nnn} in  the state $\ket{0}$ but to get the solution $A^\prime(k)=-A(k)$  we should take the expectation value of the commutator \eqref{aaa} in  the state $\ket{1}$ instead, because $\bra{1}:L_0:\ket{1}$ and  $\bra{0}::L_0::\ket{0}$ are ill-defined, an aspect that remained rather obscure in \eqref{A}. 
\par
We consider the Ward identity \cite{green}
\begin{equation}\label{ward}
\partial_-\bra{0}T (T_{++}(\sigma)T_{++}(0))\ket{0}=\frac{1}{2}\delta(\tau)\bra{0}[T_{++}(\sigma),T_{++}(0)]\ket{0}
\end{equation}
If we replace $\sigma^\alpha$ by $-\sigma^\alpha$, the derivative on the L.H.S. and hence the anomaly will change sign. It interchanges the roles of the creation and annihilation operators of the fermion and hence the roles of $\ket{0}$ and $\ket{1}$. It doesn't alter the canonical commutation relation $\{b_k,b^\dagger_k\}=1$. But this is not so for bosons.
\par
The L.H.S. of \eqref{ward} will include a factor of the form 
\begin{equation}
\int_{-\infty}^\infty dk \frac{e^{ik\tau}}{k-\omega-i\epsilon}
\end{equation}
If we replace $\sigma^\alpha$ by $-\sigma^\alpha$, $\epsilon$ will be replaced by $-\epsilon$ in the path integral. So the contour in the complex $k$-plane to evaluate the integral now should be closed in the reverse direction to yield a negative sign.
\par
From CFT \cite{ginsparg}
\begin{equation}
[L_m,L_n]_{anomalous}=\frac{c}{12}(m^3-m)\oint\frac{d\omega}{2\pi i}\omega^{m+n-1}
\end{equation}
where $\omega=e^{\tau+i\sigma}$. If we replace $\sigma^\alpha$ by $-\sigma^\alpha$, the contour will be described in the opposite direction changing the sign of the anomaly.
\par
A super-string model includes both bosonic and fermionic degrees of freedom. So unlike the previous case  not only the sign but also the magnitude of the central charge for the full theory will
 change if instead of normal order we anti-normal order the contribution of the physical fermions to the Virasoro algebra and normal order the rest. It has the potential for changing the critical dimension of a string to a more
suitable value. When the anomaly cancels we can quantize the fields successfully and define the canonical commutation relations without ambiguity to write $::L_0::$ as $:L_0:-2a$, where $a$ is the normal ordering constant, to get back to the anomaly free conventional form of the algebra that is more appropriate for low lying states.

\section{Conclusion}
\par
We studied the two dimensional U(1) current algebra and the Virasoro algebra to see that the Schwinger term is a property of the states to which the operator is applied:
if the positive k states were filled, rather than the negative k ones, the Schwinger term
has opposite sign. This corresponds to anti-normal ordering of the current-current commutator. 
\par

It remains to be seen in detail how the change of sign of  the Virasoro anomaly alter the condition of its cancellation in a consistent string theory.

\section*{Acknowledgments}
\par
I thank G.Bhattacharya for his valuable suggestions and comments.

 \end{document}